\documentclass[12pt,letterpaper]{article}
\usepackage{graphicx}
\usepackage{times}
\usepackage{mathptmx}
\begin{document}
\title{Efficient Data Assimilation for Spatiotemporal Chaos: a Local
Ensemble Transform Kalman Filter}
\date{12 December 2006}
\author{
Brian R. Hunt \\ 
\normalsize Institute for Physical Science and Technology and
Department of Mathematics \\
\normalsize University of Maryland, College Park MD 20742 \\[0.125in]
Eric J. Kostelich \\
\normalsize Department of Mathematics and Statistics \\
\normalsize Arizona State University, Tempe AZ 85287-1804 \\[0.125in]
Istvan Szunyogh \\
\normalsize Institute for Physical Science and Technology and
Department of Atmospheric and Oceanic Science \\
\normalsize University of Maryland, College Park MD 20742\\[0.125in]
}
\maketitle
\begin{abstract}

Data assimilation is an iterative approach to the problem of
estimating the state of a dynamical system using both current and past
observations of the system together with a model for the system's time
evolution.  Rather than solving the problem from scratch each time new
observations become available, one uses the model to ``forecast'' the
current state, using a prior state estimate (which incorporates
information from past data) as the initial condition, then uses
current data to correct the prior forecast to a current state
estimate.  This Bayesian approach is most effective when the
uncertainty in both the observations and in the state estimate, as it
evolves over time, are accurately quantified.  In this article, we
describe a practical method for data assimilation in large,
spatiotemporally chaotic systems.  The method is a type of ``ensemble
Kalman filter'', in which the state estimate and its approximate
uncertainty are represented at any given time by an ensemble of system
states.  We discuss both the mathematical basis of this approach and
its implementation; our primary emphasis is on ease of use and
computational speed rather than improving accuracy over previously
published approaches to ensemble Kalman filtering.  We include some
numerical results demonstrating the efficiency and accuracy of our
implementation for assimilating real atmospheric data with the global
forecast model used by the U.S. National Weather Service.

\end{abstract}

\section{Introduction}

Forecasting a physical system generally requires both a model for the
time evolution of the system and an estimate of the current state of
the system.  In some applications, the state of the system can be
measured directly with high accuracy.  In other applications, such as
weather forecasting, direct measurement of the global system state is not
feasible.  Instead, the state must be inferred from available data.
While a reasonable state estimate based on current data may be
possible, in general one can obtain a better estimate by using both
current and past data.  ``Data assimilation'' 
provides such an estimate on an ongoing basis, iteratively alternating
between a forecast step and a state estimation step; the latter step
is often called the ``analysis''.  The analysis step combines
information from current data and from a prior short-term forecast
(which is based on past data), producing a current state estimate.
This estimate is used to initialize the next short-term forecast,
which is subsequently used in the next analysis, and so on.  The data
assimilation procedure is itself a dynamical system driven by the
physical system, and the practical problem is to achieve good
``synchronization'' \cite{PC} between the two systems.

Data assimilation is widely used to study and forecast geophysical
systems \cite{GM,Kaln}.  The analysis step is generally a statistical
procedure (specifically, a Bayesian maximum likelihood estimate)
involving a prior (or ``background'') estimate of the current state
based on past data, and current data (or ``observations'') that are
used to improve the state estimate.  This procedure requires
quantification of the uncertainty in both the background state and the
observations.  While quantifying the observation uncertainty can be a
nontrivial problem, in this article we consider that problem to be
solved, and instead concentrate on the problem of quantifying the
background uncertainty.

There are two main factors that create background uncertainty.  One is
the uncertainty in the initial conditions from the previous analysis,
which produces the background state via a short-term forecast.  The
other is ``model error'', the unknown discrepancy between the model
dynamics and actual system dynamics.  Quantifying the uncertainty due
to model error is a challenging problem, and while this problem
generally cannot be ignored in practice, we discuss only crude ways
of accounting for it in this article.  For the time being, let us
consider an idealized ``perfect model'' scenario, in which there is no
model error.

The main purpose of this article is to describe a practical framework
for data assimilation that 
is both relatively easy to implement and computationally efficient,
even for large, spatiotemporally chaotic systems.  (By
``spatiotemporally chaotic'' we mean a spatially extended system that
exhibits temporally chaotic behavior with weak long-range spatial
correlations.)  The emphasis here is on methodology that scales well
to high-dimensional systems and large numbers of observations, rather
than on what would be optimal given unlimited computational resources.
Ideally, one would keep track of a probability distribution of system
states, propagating the distribution using the
Fokker-Planck-Kolmogorov equation during the forecast step.  While
this approach provides a theoretical basis for the methods used in
practice \cite{Jaz}, it would be computationally expensive even for a
low-dimensional system and is not at all feasible for a high-dimensional
system.  Instead one can use a Monte Carlo approach, using a large
ensemble of system states to approximate the distribution (see
\cite{DFG} for an overview), or a parametric approach like the Kalman
filter \cite{Kalm,KB}, which assumes Gaussian distributions and tracks
their mean and covariance.  (The latter approach was derived
originally for linear problems, but serves as a reasonable
approximation for nonlinear problems when the uncertainties remain
sufficiently small.)

The methodology of this article is based on the Ensemble Kalman Filter
\cite{Eve1,Eve2,Eve3}, which has elements of both approaches: it uses the
Gaussian approximation and follows the time evolution of the mean and
covariance by propagating an ensemble of states.  The ensemble can be
reasonably small relative to other Monte Carlo methods because it is
used only to parametrize the distribution, not to sample it
thoroughly.  The ensemble should be large enough to approximately
span the space of possible system states at a given time, because the
analysis essentially determines which linear combination of the
ensemble members forms the best estimate of the current state, given
the current observations. 

Many variations on the Ensemble Kalman Filter have been published in
the geophysical literature, and this article draws ideas from a number
of them \cite{And,AA,BEM,HWS,HM1,HM2,Kep,Ott1,Ott2,WH,Zup}.  These
articles in turn draw ideas both from earlier work on geophysical data
assimilation and from the engineering and mathematics literature on
nonlinear filtering.  For the most part, we limit our citations to
ensemble-based articles rather than attempt to trace all ideas to
their original sources.  We call the method described here a Local
Ensemble Transform Kalman Filter (LETKF), because it is most closely
related to the Local Ensemble Kalman Filter \cite{Ott1,Ott2}
and the Ensemble Transform Kalman Filter \cite{BEM}.  Indeed, it can
produce analyses that are equivalent to the LEKF in a more efficient
manner that is formally similar to the ETKF.  While this article
does not describe a fundamentally new method for data assimilation,
it proposes a significant refinement of previously published
approaches that combines formal simplicity with the flexibility to
adapt to a variety of applications.

In Section 2, we start by posing a general problem about which
trajectory of a dynamical system ``best fits'' a time series of data;
this problem is solved exactly for linear problems by the Kalman
filter and approximately for nonlinear problems by ensemble Kalman
filters.  Next we derive the Kalman filter equations as a guide for
what follows.  Then we discuss ensemble Kalman filters in general and
the issue of ``localization'', which is important for applications to
spatiotemporally chaotic systems.  Finally, we develop the basic LETKF
equations, which provide a framework for data assimilation
that allows a system-dependent localization strategy to be developed
and tuned.  We discuss also several options for ``covariance
inflation'' to compensate for the effects of model error and the
deficiencies due to small sample size and linear approximation that
are inherent to ensemble Kalman filters.

In Section 3, we give step-by-step instructions for efficient
implementation of the approach developed in Section 2 and discuss
options for further improving computational speed in certain cases.
Then in Section 4, we present a generalization that allows observations
gathered at different times to be assimilated simultaneously in a
natural way.  In Section 5, we present preliminary results using a
global atmospheric forecast model with real observations; these
results compare favorably with the data assimilation method used by
the National Weather Service, and demonstrate the feasibility of the
LETKF algorithm for large models and data sets.  Section 6 is a brief
conclusion.  The notation in this article is based largely on that
proposed in \cite{ICGL}, with some elements from \cite{Ott2}.

\section{Mathematical Formulation}

Consider a system governed by the ordinary differential equation
\begin{equation} \label{ode}
\frac{d{\bf x}}{dt} = F(t,{\bf x}),
\end{equation}
where ${\bf x}$ is an $m$-dimensional vector representing the state of
the system at a given time.  Suppose we are given a set of (noisy)
observations of the system made at various times, and we want to
determine which trajectory $\{{\bf x}(t)\}$ of (\ref{ode}) ``best''
fits the observations.  For any given $t$, this trajectory gives an
estimate of the system state at time $t$.

To formulate this problem mathematically, we need to define ``best
fit'' in this context.  Let us assume that the observations are the
result of measuring quantities that depend on the system state in a
known way, with Gaussian measurement errors.  In other words, an
observation at time $t_j$ is a triple $({\bf y}^o_j, H_j, {\bf R}_j)$,
where ${\bf y}^o_j$ is a vector of observed values, and $H_j$ and ${\bf
R}_j$ describe the relationship between ${\bf y}^o_j$ and ${\bf
x}(t_j)$:
$$
{\bf y}^o_j = H_j({\bf x}(t_j)) + {\bf \varepsilon}_j,
$$
where ${\bf \varepsilon}_j$ is a Gaussian random variable with mean
${\bf 0}$ and covariance matrix ${\bf R}_j$.  Notice that we are
assuming a perfect model here: the observations are based on a
trajectory of $(\ref{ode})$, and our problem is simply to infer which
trajectory produced the observations.  In a real application, the
observations come from a trajectory of the physical system for which
$(\ref{ode})$ is only a model.  So a more realistic (but more
complicated) problem would be to determine a pseudo-trajectory of
$(\ref{ode})$, or a trajectory of an associated stochastic
differential equation, that best fits the observations.  Formulating this
problem mathematically then requires some assumptions about the size and
nature of the model error.  We use the perfect model problem as
motivation and defer the consideration of model error until later.

Given our assumptions about the observations, we can formulate a maximum
likelihood estimate for the trajectory of (\ref{ode}) that best fits
the observations at times $t_1 < t_2 < \cdots < t_n$.  The likelihood of
a trajectory ${\bf x}(t)$ is proportional to
$$
\prod_{j=1}^n \exp\left(-\frac{1}{2}[{\bf y}^o_j - H_j({\bf x}(t_j))]^T {\bf
R}_j^{-1} [{\bf y}^o_j - H_j({\bf x}(t_j))]\right).
$$
The most likely trajectory is the one that maximizes this expression,
or equivalently minimizes the ``cost function''
\begin{equation} \label{trajcost}
J^o(\{{\bf x}(t)\}) = \sum_{j=1}^n [{\bf y}^o_j - H_j({\bf x}(t_j))]^T
{\bf R}_j^{-1} [{\bf y}^o_j - H_j({\bf x}(t_j))].
\end{equation}
Thus, the ``most likely'' trajectory is also the one that best fits the
observations in a least square sense.

Notice that (\ref{trajcost}) expresses the cost $J^o$ as a function of
the trajectory $\{{\bf x}(t)\}$.  To minimize the cost, it is more
convenient to write $J^o$ as a function of the system state at a
particular time $t$.  Let $M_{t,t'}$ be the map that propagates a
solution of (\ref{ode}) from time $t$ to time $t'$.\footnote{In the
derivations that follow, we allow $t'$ to be less than $t$, though in
practice integrating (\ref{ode}) backward in time may be problematic
--- for example, if (\ref{ode}) represents a discretization of a
dissipative partial differential equation.  Our use of $M_{t,t'}$ for
$t' < t$ is entirely expository; the methodology we develop will not
require backward integration of (\ref{ode}).}  Then
\begin{equation} \label{statecost}
J^o_t({\bf x}) = \sum_{j=1}^n [{\bf y}^o_j - H_j(M_{t,t_j}({\bf x}))]^T
{\bf R}_j^{-1} [{\bf y}^o_j - H_j(M_{t,t_j}({\bf x}))]
\end{equation}
expresses the cost in terms of the system state ${\bf x}$ at time $t$.
Thus to estimate the state at time $t$, we attempt to minimize $J^o_t$.

For a nonlinear model, there is no guarantee that a unique minimizer
exists.  And even if it does, evaluating $J^o_t$ is apt to be
computationally expensive, and minimizing it may be impractical.  But
if both the model and the observation operators $H_j$ are linear, the
minimization is quite tractable, because $J^o_t$ is then quadratic.
Furthermore, instead of performing the minimization from scratch at
each successive time $t_n$, one can compute the minimizer by an
iterative method, namely the Kalman filter \cite{Kalm,KB}, which we
now describe in the perfect model scenario.  This method forms
the basis for the approach we will use in the nonlinear scenario.

\subsection{Linear Scenario: the Kalman Filter}
In the linear scenario, we can write $M_{t,t'}({\bf x}) = {\bf
M}_{t,t'} {\bf x}$ and $H_j({\bf x}) = {\bf H}_j {\bf x}$ where ${\bf
M}_{t,t'}$ and ${\bf H}_j$ are matrices.
Using the terminology from the introduction, we now describe how to
perform a forecast step from time $t_{n-1}$ to time $t_n$ followed by
an analysis step at time $t_n$, in such a way that if we start with the
most likely system state, in the sense described above, given the
observations up to time $t_{n-1}$, we end up with the most likely
state given the observations up to time $t_n$.  The forecast step
propagates the solution from time $t_{n-1}$ to time $t_n$, and the
analysis step combines the information provided by the observations at
time $t_n$ with the propagated information from the prior observations.
This iterative approach requires that we keep track of not only the most
likely state, but also its uncertainty, in the sense described below.
(Of course, the fact that the Kalman filter computes the uncertainty
in its state estimate may be viewed as a virtue.)

Suppose the analysis at time $t_{n-1}$ has produced a state estimate
$\bar{\bf x}^a_{n-1}$ and an associated covariance matrix ${\bf
P}^a_{n-1}$.  In probabilistic terms, $\bar{\bf x}^a_{n-1}$ and
${\bf P}^a_{n-1}$ represent the mean and covariance of a Gaussian
probability distribution that represents the relative likelihood of
the possible system states given the observations from time $t_1$
to $t_{n-1}$.  Algebraically, what we assume is that for some constant
$c$,
\begin{equation} \label{gaussian1}
\sum_{j=1}^{n-1} [{\bf y}^o_j - {\bf H}_j {\bf M}_{t_{n-1},t_j}{\bf
x}]^T {\bf R}_j^{-1} [{\bf y}^o_j - {\bf H}_j {\bf M}_{t_{n-1},t_j}{\bf
x}] = [{\bf x} - \bar{\bf x}^a_{n-1}]^T ({\bf P}^a_{n-1})^{-1} [{\bf
x} - \bar{\bf x}^a_{n-1}] + c.
\end{equation}
In other words, the analysis at time $t_{n-1}$ has ``completed the
square'' to express the part of the quadratic cost function
$J^o_{t_{n-1}}$ that depends on the observations up to that time as a
single quadratic form plus a constant.  The Kalman filter determines
$\bar{\bf x}^a_n$ and ${\bf P}^a_n$ such that an analogous equation holds
at time $t_n$.

First we propagate the analysis state estimate $\bar{\bf x}^a_{n-1}$
and its covariance ${\bf P}^a_{n-1}$ using the forecast model
to produce a background state estimate $\bar{\bf x}^b_n$ and
covariance matrix ${\bf P}^b_n$ for the next analysis:
\begin{equation} \label{prop1}
\bar{\bf x}^b_n = {\bf M}_{t_{n-1},t_n} \bar{\bf x}^a_{n-1}, \hspace{42pt}
\end{equation}
\begin{equation} \label{prop2}
{\bf P}^b_n = {\bf M}_{t_{n-1},t_n} {\bf P}^a_{n-1}
{\bf M}_{t_{n-1},t_n}^T.
\end{equation}
Under a linear model, a Gaussian distribution of states at one time
propagates to a Gaussian distribution at any other time, and the
equations above describe how the model propagates the mean and
covariance of such a distribution.  (Usually, the Kalman filter adds a
constant matrix to the right side of (\ref{prop2}) to represent
additional uncertainty due to model error.)

Next, we want to rewrite the cost function $J^o_{t_n}$ given by
(\ref{statecost}) in terms
of the background state estimate and the observations at time $t_n$.
(This step is often formulated as applying Bayes' rule to the
corresponding probability density functions.)
In (\ref{gaussian1}), ${\bf x}$ represents a hypothetical system state
at time $t_{n-1}$.  In our expression for $J^o_{t_n}$, we want ${\bf x}$
to represent instead a hypothetical system state at time $t_n$, so we
first replace ${\bf x}$ by ${\bf M}_{t_n,t_{n-1}} {\bf x} = {\bf
M}_{t_{n-1},t_n}^{-1} {\bf x}$ in (\ref{gaussian1}).  Then using
(\ref{prop1}) and (\ref{prop2}) yields
$$
\sum_{j=1}^{n-1} [{\bf y}^o_j - {\bf H}_j {\bf M}_{t_n,t_j}{\bf x}]^T
{\bf R}_j^{-1} [{\bf y}^o_j - {\bf H}_j {\bf M}_{t_n,t_j}{\bf x}] =
[{\bf x} - \bar{\bf x}^b_n]^T ({\bf P}^b_n)^{-1} [{\bf x} - {\bf
x}^b_n] + c.
$$
It follows that
\begin{equation} \label{linkalcost}
J^o_{t_n}({\bf x}) = [{\bf x} - \bar{\bf x}^b_n]^T ({\bf P}^b_n)^{-1}
[{\bf x} - \bar{\bf x}^b_n] + [{\bf y}^o_n - {\bf H}_n {\bf x}]^T {\bf
R}_n^{-1} [{\bf y}^o_n - {\bf H}_n {\bf x}] + c.
\end{equation}

To complete the data assimilation cycle, we determine the state
estimate $\bar{\bf x}^a_n$ and its covariance ${\bf P}^a_n$ so that
$$
J^o_{t_n}({\bf x}) = [{\bf x} - \bar{\bf x}^a_n]^T ({\bf P}^a_n)^{-1}
[{\bf x} - \bar{\bf x}^a_n] + c'
$$
for some constant $c'$.  Equating the terms of degree $2$ in ${\bf
x}$, we get
\begin{equation} \label{ancov}
{\bf P}^a_n = \left[({\bf P}^b_n)^{-1} + {\bf H}_n^T {\bf R}_n^{-1}
{\bf H}_n\right]^{-1}.
\end{equation}
Equating the terms of degree $1$, we get
\begin{equation} \label{anmean}
\bar{\bf x}^a_n = {\bf P}^a_n \left[({\bf P}^b_n)^{-1} \bar{\bf x}^b_n + {\bf
H}_n^T {\bf R}_n^{-1} {\bf y}^o_n\right].
\end{equation}
Notice that when the model state is observed directly, ${\bf H}_n$ is
the identity matrix, and equation (\ref{anmean}) expresses the
analysis state estimate as a weighted average of the background state
estimate and the observations, weighted according to the inverse
covariance of each.

Equations (\ref{ancov}) and (\ref{anmean}) can be written in many
different but equivalent forms, and it will be useful later to rewrite
both of them now.  Using (\ref{ancov}) to eliminate $({\bf
P}^b_n)^{-1}$ from (\ref{anmean}) yields
\begin{equation} \label{anmean'}
\bar{\bf x}^a_n = \bar{\bf x}^b_n + {\bf P}^a_n {\bf H}_n^T {\bf R}_n^{-1}
({\bf y}^o_n - {\bf H}_n \bar{\bf x}^b_n).
\end{equation}
The matrix ${\bf P}^a_n {\bf H}_n^T {\bf R}_n^{-1}$ is called the
``Kalman gain''.  It multiplies the difference between the
observations at time $t_n$ and the values predicted by the background
state estimate to yield the increment between the background and
analysis state estimates.  Next, multiplying (\ref{ancov}) on the
right by $({\bf P}^b_n)^{-1} {\bf P}^b_n$ and combining the inverses
yields
\begin{equation} \label{ancov'}
{\bf P}^a_n = ({\bf I} + {\bf P}^b_n {\bf H}_n^T {\bf R}_n^{-1} {\bf
H}_n)^{-1} {\bf P}^b_n.
\end{equation}
This expression provides a more efficient way than (\ref{ancov}) to
compute ${\bf P}^a_n$, since it does not require inverting ${\bf
P}^b_n$.

\paragraph{Initialization.}
The derivation above of the Kalman filter avoids the issue of how to
initialize the iteration.  To solve the best fit problem we
originally posed, we should make no assumptions about the system state
prior to the analysis at time $t_1$.  Formally we can regard the
background covariance ${\bf P}^b_1$ to be infinite, and for $n = 1$
use (\ref{ancov}) and (\ref{anmean}) with $({\bf P}^b_1)^{-1} = {\bf
0}$.  This works if there are enough observations at time $t_1$ to
determine (aside from the measurement errors) the system state; that
is, if ${\bf H}_1$ has rank equal to the number of model variables
$m$.  The analysis then determines $\bar{\bf x}^a_1$ in the appropriate
least-square sense.  However, if there are not enough observations,
then the matrix to be inverted in (\ref{ancov}) does not have full
rank.  To avoid this difficulty, one can assume a prior background
distribution at time $t_1$, with ${\bf P}^b_1$ reasonably large but
finite.  This adds a small quadratic term to the cost function being
minimized, but with sufficient observations over time, the effect of
this term on the analysis at time $t_n$ decreases in significance as
$n$ increases.

\subsection{Nonlinear Scenario: Ensemble Kalman Filtering} \label{nonlin}

Many approaches to data assimilation for nonlinear problems are based
on the Kalman filter, or at least on minimizing a cost function
similar to (\ref{linkalcost}).  At a minimum, a nonlinear model forces a
change in the forecast equations (\ref{prop1}) and (\ref{prop2}),
while nonlinear observation operators $H_n$ force a change in the
analysis equations (\ref{anmean'}) and (\ref{ancov'}).  The extended
Kalman filter (see, for example, \cite{Jaz}) computes $\bar{\bf x}^b_n =
M_{t_{n-1},t_n}(\bar{\bf x}^a_{n-1})$ using the nonlinear model, but
computes ${\bf P}^b_n$ using the linearization ${\bf M}_{t_{n-1},t_n}$
of $M_{t_{n-1},t_n}$ around $\bar{\bf x}^a_{n-1}$.  The analysis then uses
the linearization ${\bf H}_n$ of $H_n$ around $\bar{\bf x}^b_n$.  This
approach is problematic for complex, high-dimensional models such as a
global weather model for (at least) two reasons.  First, it is not
easy to linearize such a model.  Second, when the number of model
variables $m$ is several million, computations involving the $m \times
m$ covariance matrices are very expensive.

Approaches used in operational weather forecasting generally eliminate,
for pragmatic reasons, the time iteration of the Kalman filter.  The
U.S. National Weather Service
performs data assimilation every 6 hours using the ``3D-Var'' method
\cite{Lorenc1,PD}, in which the background covariance ${\bf P}^b_n$ in
(\ref{linkalcost}) is replaced by a constant matrix ${\bf B}$ representing
typical uncertainty in a 6-hour forecast.  This simplification allows
the analysis to be formulated in a manner that precomputes the most
expensive matrix operations, so that they do not have to be repeated at
each time $t_n$.  The 3D-Var cost function also
allows a nonlinear observation operator $H_n$, and is minimized
numerically to produce the analysis state estimate ${\bf x}^a_n$.

The ``4D-Var'' method \cite{LT,TC} used by the European
Centre for Medium-Range Weather Forecasts uses a cost function that
includes a constant-covariance background term as in 3D-Var, together
with a sum like (\ref{trajcost}) accounting for the observations collected
over a 12-hour time window.  Again the cost function is minimized
numerically; this procedure is computationally intensive, because 
computing the gradient of the 4D-Var cost function requires
integrating both the nonlinear model and its linearization over the
12-hour window, and this procedure is repeated until a satisfactory
approximation to the minimum is found.

The key idea of ensemble Kalman filtering \cite{Eve1,Eve3} is to choose
at time $t_{n-1}$ an ensemble of initial conditions whose spread
around $\bar{\bf x}^a_{n-1}$ characterizes the analysis covariance ${\bf
P}^a_{n-1}$, propagate each ensemble member using the nonlinear model,
and compute ${\bf P}^b_n$ based on the resulting ensemble at time
$t_n$.  Thus like the extended Kalman filter, the (approximate)
uncertainty in the state estimate is propagated from one analysis to
the next, unlike 3D-Var (which does not propagate the uncertainty at
all) or 4D-Var (which propagates it only with the time window over
which the cost function is minimized).  Furthermore, ensemble Kalman
filters do this without requiring a linearized model.  On the other
hand, 4D-Var (with ``weak constraint'') allows a wide variety of model
error terms to be incorporated into the cost function.

In spite of their differences, though, we emphasize that in the
absence of computational limitations, 4D-Var and ensemble Kalman
filtering should be able to produce similar results because they both
seek to minimize the same type of cost function.  Indeed, in a perfect
model scenario \cite{FHH}, we obtained similar results with both
methods when we used a sufficiently long time window for 4D-Var and
when we used enough ensemble members and performed the analysis
sufficiently frequently in a 4D version (described in
Section~\ref{4dletkf} of this article) of our LETKF.  In atmospheric
data assimilation, ensemble Kalman filtering has not yet equaled the
best results using 4D-Var, but it has begun to achieve results that
compare favorably with operational 3D-Var results
\cite{HMPBCSH,WHWST,MY}; see also Section~\ref{results}.

Perhaps the most important difference between ensemble Kalman
filtering and the other methods described above is that the former quantifies
uncertainty only in the space spanned by the ensemble.  Assuming
that computational resources restrict the number of ensemble members
$k$ to be much smaller than the number of model variables $m$, this
can be a severe limitation.  On the other hand, if this limitation can
be overcome (see the section on ``Localization'' below), then the
analysis can be performed in a much lower-dimensional space ($k$
versus $m$).  Thus, ensemble Kalman filtering has the potential to be
more computationally efficient than the other methods.  Indeed, the
main point of this article is to describe how to do ensemble Kalman
filtering efficiently without sacrificing accuracy.

\paragraph{Notation.}
We start with an ensemble $\{{\bf x}^{a(i)}_{n-1}: i = 1,2,\ldots,k\}$
of $m$-dimensional model state vectors at time $t_{n-1}$.  One
approach would be to let one of the ensemble members represent the
best estimate of the system state, but here we assume the ensemble to
be chosen so that its average represents the analysis state estimate.
We evolve each ensemble member according to the nonlinear model to
obtain a background ensemble $\{{\bf x}^{b(i)}_n: i = 1,2,\ldots,k\}$
at time $t_n$:
$$
{\bf x}^{b(i)}_n = M_{t_{n-1},t_n}({\bf x}^{a(i)}_{n-1}).
$$
For the rest of this article, we will discuss what to do at the analysis
time $t_n$, and so we now drop the subscript $n$.  Thus, for example,
$H$ and ${\bf R}$ will represent respectively the observation operator
and the observation error covariance matrix at the analysis time.  Let
$\ell$ be the number of scalar observations used in the analysis.

For the background state estimate and its covariance, we use the sample
mean and covariance of the background ensemble:
$$
\bar{\bf x}^b = k^{-1} \sum_{i=1}^k {\bf x}^{b(i)},
$$
\begin{equation} \label{bgcov}
{\bf P}^b = (k-1)^{-1} \sum_{i=1}^k ({\bf x}^{b(i)} - \bar{\bf x}^b)
({\bf x}^{b(i)} - \bar{\bf x}^b)^T = (k-1)^{-1} {\bf X}^b ({\bf
X}^b)^T,
\end{equation}
where ${\bf X}^b$ is the $m \times k$ matrix whose $i$th column is
${\bf x}^{b(i)} - \bar{\bf x}^b$.  (Notice that the rank of ${\bf
P}^b$ is equal to the rank of ${\bf X}^b$, which is at most $k - 1$
because the sum of its columns is ${\bf 0}$.  Thus, the ensemble size
limits the rank of the background covariance matrix.)
The analysis must determine not only an state estimate $\bar{\bf x}^a$
and covariance ${\bf P}^a$, but also an ensemble $\{{\bf x}^{a(i)}: i =
1,2,\ldots,k\}$ with the appropriate sample mean and covariance:
$$
\bar{\bf x}^a = k^{-1} \sum_{i=1}^k {\bf x}^{a(i)},
$$
\begin{equation} \label{enscov}
{\bf P}^a = (k-1)^{-1} \sum_{i=1}^k ({\bf x}^{a(i)} - \bar{\bf x}^a)
({\bf x}^{a(i)} - \bar{\bf x}^a)^T = (k-1)^{-1} {\bf X}^a ({\bf
X}^a)^T,
\end{equation}
where ${\bf X}^a$ is the $m \times k$ matrix whose $i$th column is
${\bf x}^{a(i)} - \bar{\bf x}^a$.

In Section~\ref{letkf}, we will describe how to determine $\bar{\bf
x}^a$ and ${\bf P}^a$ for a (possibly) nonlinear observation operator
$H$ in a way that agrees with the Kalman filter equations
(\ref{anmean'}) and (\ref{ancov'}) in the case that $H$ is linear.

\paragraph{Choice of analysis ensemble.}
Once $\bar{\bf x}^a$ and ${\bf P}^a$ are specified, there are still many
possible choices of an analysis ensemble (or equivalently, a matrix
${\bf X}^a$ that satisfies (\ref{enscov}) and the sum of whose columns
is zero).  Many ensemble Kalman filters have been proposed,
and one of the main differences among them is how the analysis
ensemble is chosen.  The simplest approach is to apply the Kalman
filter update (\ref{anmean'}) separately to each background ensemble
member (rather than their mean) to get the corresponding analysis
ensemble member.  However, this results in an analysis ensemble whose
sample covariance is smaller than the analysis covariance ${\bf P}^a$
given by (\ref{ancov'}), unless the observations
are artificially perturbed so that each ensemble member is updated
using different random realization of the perturbed observations
\cite{BVE,HM1}.  Ensemble square-root filters
\cite{And,WH,BEM,TABHW,Ott1,Ott2} instead use more involved but
deterministic algorithms to generate an analysis ensemble with the
desired sample mean and covariance.  As such, their analyses coincide
exactly with the Kalman filter equations in the linear scenario of the
previous section.  We will use this deterministic approach in
Section~\ref{letkf}.

\paragraph{Localization.}
Another important issue in ensemble Kalman filtering of
spatiotemporally chaotic systems is spatial localization.  If the
ensemble has $k$ members, then the background covariance matrix ${\bf
P}^b$ given by (\ref{bgcov}) describes nonzero uncertainty only in the
(at most) $k$-dimensional subspace spanned by the ensemble, and a global
analysis will allow adjustments to the system state only in this
subspace.  If the system is high-dimensionally unstable --- more
precisely, if it has more than $k$ positive Lyapunov exponents ---
then forecast
errors will grow in directions not accounted for by the ensemble, and
these errors will not be corrected by the analysis.  On the other
hand, in a sufficiently small local region, the system may behave like
a low-dimensionally unstable system driven by the dynamics in
neighboring regions; such behavior was observed for a global weather
model in \cite{PHKYO,OSP}.  Performing the analysis locally requires
the ensemble to represent uncertainty in only the local unstable space.  By
allowing the local analyses to choose different linear combinations of
the ensemble members in different regions, the global analysis is not
confined to the $k$-dimensional ensemble space and instead explores a
much higher-dimensional space \cite{Fuk,Ott1,Ott2}.  Another
explanation for the
necessity of localization for spatiotemporally chaotic systems is that the
limited sample size provided by an ensemble will produce spurious
correlations between distant locations in the background covariance
matrix ${\bf P}^b$ \cite{HM1,HWS}.  Unless they are suppressed, these
spurious correlations will cause observations from one location to
affect, in an essentially random manner, the analysis in locations an
arbitrarily
large distance away.  If the system has a characteristic ``correlation
distance'', then the analysis should ignore ensemble correlations over
much larger distances.  In addition to providing better results in
many cases, localization allows the analysis to be done more
efficiently as a parallel computation \cite{Kep,Ott1,Ott2}.

Localization is generally done either explicitly, considering only the
observations from a region surrounding the location of the analysis
\cite{KT,HM1,Kep,And,Ott1,Ott2}, or implicitly, by multiplying the
entries in ${\bf P}^b$ by a distance-dependent function that decays to
zero beyond a certain distance, so that observations do not affect the
model state beyond that distance \cite{HM2,HWS,WH}.  We will follow the
explicit approach here, doing a separate analysis for each spatial
grid point of the model.  (Our use of ``grid point'' assumes the model
to be a discretization of a partial differential equation, or
otherwise to be defined on a lattice, but the method is also applicable
to systems with other geometries.)  The choice of which observations
to use for each grid point is up to the user of the method, and a good
choice will depend both on the particular system being modeled and on the
size of the ensemble (more ensemble members generally allow more
distant observations to be used gainfully).  It is important, however,
that most of the observations used in the analysis at a particular
grid point also be used in the analysis at neighboring grid points.
This ensures that the analysis ensemble does not change suddenly from
one grid point to the next.  For an atmospheric model, a reasonable
approach is to use observations within a cylinder of a given radius
and height centered at the analysis grid point and to determine
empirically which values of the radius and height
work best.  At its simplest, the method we describe gives all of the
chosen observations equal influence on the analysis, but we will
also describe how to make their influence decay gradually toward zero
as their distance from the analysis location increases.

\paragraph{Initial background ensemble.}
A common method for generating a background ensemble to use at the
first analysis time is to run the model for a while and to select
model states at different randomly chosen times.  The intent of this
method is for the initial background ensemble to be sampled from a
climatological distribution.  This is a reasonable choice for the
background distribution when no prior observations are available.

\subsection{LETKF: A Local Ensemble Transform Kalman Filter} \label{letkf}

We now describe an efficient means of performing the analysis that
transforms a background ensemble $\{{\bf x}^{b(i)}: i =
1,2,\ldots,k\}$ into an appropriate analysis ensemble $\{{\bf
x}^{a(i)}: i = 1,2,\ldots,k\}$, using the notation defined above.  We
assume that the number of ensemble members $k$ is smaller than both
the number of model variables $m$ and the number of observations
$\ell$,\footnote{This assumption is only expository; our algorithm does
not require an upper bound on $k$, but it is less efficient than doing
the analysis in model space if $m < k$ or in observation space if $\ell
< k$.} even when localization has reduced the effective values of $m$
and $\ell$ considerably compared to a global analysis.  (In this
section we assume that the choice of observations to use for the local
analysis has been performed already, and consider ${\bf y}^o$, $H$,
and ${\bf R}$ to be truncated to these observations; as such,
correlations between errors in the chosen observations and errors in
other observations are ignored.)  Most of the analysis takes place
in a $k$-dimensional space, with as few operations as possible in the
model and observation spaces.

Formally, we want the analysis mean $\bar{\bf x}^a$ to minimize
the Kalman filter cost function (\ref{linkalcost}), modified to allow
for a nonlinear observation operator $H$:
\begin{equation} \label{kalcost}
J({\bf x}) = ({\bf x} - \bar{\bf x}^b)^T ({\bf P}^b)^{-1} ({\bf x} -
\bar{\bf x}^b) + [{\bf y}^o - H({\bf x})]^T {\bf R}^{-1} [{\bf y}^o -
H({\bf x})].
\end{equation}
However, the $m \times m$ background covariance matrix ${\bf P}^b =
(k-1)^{-1} {\bf X}^b ({\bf X}^b)^T$ has rank at most $k - 1$, and
is therefore not invertible.  Nonetheless, as a symmetric matrix, it is
one-to-one on its column space $S$, which is also the column space of
${\bf X}^b$, or in other words the space spanned by the background
ensemble perturbations.  So in this sense, $({\bf P}^b)^{-1}$ is
well-defined on $S$.  Then
$J$ is also well-defined
for ${\bf x} - \bar{\bf x}^b$ in $S$, and the minimization can be
carried out in this subspace.\footnote{Considerably more general cost
functions can be used, at the expense of having to perform the
minimization numerically in the ensemble space $S$.  Since this space
is relatively low-dimensional, this approach is still feasible even
for high-dimensional systems.  In \cite{HH1} we use a non-quadratic
background term within the LETKF framework, and \cite{Zup}
uses a similar approach to allow a non-quadratic observation term.}
As we have said, this reduced
dimensionality is an advantage from the point of view of efficiency,
though the restriction of the analysis mean to $S$ is sure to be
detrimental if $k$ is too small.

To perform the analysis on $S$, we must choose an appropriate
coordinate system.  A natural approach is to use the singular vectors
of ${\bf X}^b$ (the eigenvectors of ${\bf P}^b$) to form a basis for
$S$ \cite{And,Ott1,Ott2}.  Here we avoid this step by using instead the
columns of ${\bf X}^b$ to span $S$, as in \cite{BEM}.  One conceptual
difficulty in this approach is that the sum of these columns is zero,
so they are necessarily linearly dependent.  We could assume the
first $k - 1$ columns to be independent and use them as a basis, but
this assumption is unnecessary and clutters the resulting equations.
Instead, we regard ${\bf X}^b$ as a linear transformation from a
$k$-dimensional space $\tilde{S}$ onto $S$, and perform the analysis
in $\tilde{S}$.  Let ${\bf w}$ denote a vector in $\tilde{S}$; then
${\bf X}^b {\bf w}$ belongs to the space $S$ spanned by the background
ensemble perturbations, and ${\bf x} = \bar{\bf x}^b + {\bf X}^b {\bf
w}$ is the corresponding model state.

Notice that if ${\bf w}$ is a Gaussian random vector with mean ${\bf
0}$ and covariance $(k - 1)^{-1} {\bf I}$, then ${\bf x} = \bar{\bf
x}^b + {\bf X}^b {\bf w}$ is Gaussian with mean $\bar{\bf
x}^b$ and covariance ${\bf P}^b = (k - 1)^{-1} {\bf X}^b ({\bf
X}^b)^T$.  This motivates the cost function
\begin{equation} \label{tildecost}
\tilde{J}({\bf w}) = (k-1) {\bf w}^T {\bf w} + [{\bf y}^o - H(\bar{\bf
x}^b + {\bf X}^b {\bf w})]^T {\bf R}^{-1} [{\bf y}^o - H(\bar{\bf x}^b
+ {\bf X}^b {\bf w})]
\end{equation}
on $\tilde{S}$.  In particular, we claim that if $\bar{\bf w}^a$
minimizes $\tilde{J}$, then $\bar{\bf x}^a = \bar{\bf x}^b + {\bf X}^b
\bar{\bf w}^a$ minimizes the cost function $J$.  Substituting the
change of variables formula into (\ref{kalcost}) and using (\ref{bgcov})
yields the identity
\begin{equation} \label{costdiff}
\tilde{J}({\bf w}) = (k-1) {\bf w}^T ({\bf I} - ({\bf X}^b)^T [{\bf
X}^b ({\bf X}^b)^T]^{-1} {\bf X}^b) {\bf w} + J(\bar{\bf x}^b + {\bf
X}^b {\bf w}).
\end{equation}
The matrix ${\bf I} - ({\bf X}^b)^T [{\bf X}^b ({\bf X}^b)^T]^{-1}
{\bf X}^b$ is the orthogonal projection onto the null space $N$ of
${\bf X}^b$.  (Generally $N$ will be one-dimensional, spanned by the
vector $(1, 1, \ldots, 1)^T$, but it could be higher-dimensional.)
Thus, the first term on the right side of (\ref{costdiff}) depends
only on the component of ${\bf w}$ in $N$, while the second term
depends only on its component in the space orthogonal to $N$ (which is
in one-to-one correspondence with $S$ under ${\bf X}^b$).  Thus if
$\bar{\bf w}^a$ minimizes $\tilde{J}$, then it must be orthogonal to
$N$, and the corresponding vector $\bar{\bf x}^a$ minimizes $J$.

A cost function equivalent to (\ref{tildecost}) appears in
\cite{Lorenc2}.  More generally, implementations of 3D-Var and 4D-Var
commonly use a preconditioning step that expresses the cost function
in a form similar to (\ref{tildecost}).

\paragraph{Nonlinear observations.}
The most accurate way to allow for a nonlinear observation operator
$H$ would be to numerically minimize $\tilde{J}$ in the
$k$-dimensional space $\tilde{S}$, as in \cite{Zup}.  If $H$ is
sufficiently nonlinear, then $\tilde{J}$ could have multiple minima,
but a numerical minimization using ${\bf w} = {\bf 0}$ (corresponding
to ${\bf x} = \bar{\bf x}^b$) as an initial guess would still be a
reasonable approach.  Having determined $\bar{\bf w}^a$ in this
manner, one would compute the analysis covariance $\tilde{\bf P}^a$ in
$\tilde{S}$ from the second partial derivatives of $\tilde{J}$ at
$\bar{\bf w}^a$, then use ${\bf X}^b$ to transform the analysis
results into the model space, as below.  But to formulate the
analysis more explicitly, we now linearize $H$ about the background
ensemble mean $\bar{\bf x}^b$.  Of course, if $H$ is linear then we
will find the minimum of $\tilde{J}$ exactly.  And if the spread of
the background ensemble is not too large, the linearization should be
a decent approximation, similar to the approximation we have already
made that a linear combination of background ensemble members is also
a plausible background model state.

Since we only need to evaluate $H$ in the ensemble space (or
equivalently to evaluate $H(\bar{\bf x}^b + {\bf X}^b {\bf w})$ for
${\bf w}$ in $\tilde{S}$), the simplest way to linearize $H$ is to
apply it to each of the ensemble members ${\bf x}^{b(i)}$ and
interpolate.  To this end, we define an ensemble ${\bf y}^{b(i)}$ of
background observation vectors by
\begin{equation} \label{obsens}
{\bf y}^{b(i)} = H({\bf x}^{b(i)}).
\end{equation}
We define also their mean $\bar{\bf y}^b$, and the $\ell \times k$ matrix
${\bf Y}^b$ whose $i$th column is ${\bf y}^{b(i)} - \bar{\bf y}^b$.
We then make the linear approximation
\begin{equation} \label{linobs}
H(\bar{\bf x}^b + {\bf X}^b {\bf w}) \approx \bar{\bf y}^b + {\bf Y}^b
{\bf w}.
\end{equation}
The same approximation is used in, for example, \cite{HM2}, and is
equivalent to the joint state-observation space method in \cite{And}.

\paragraph{Analysis.}
The linear approximation we have just made yields the quadratic cost
function
\begin{equation} \label{lincost}
\tilde{J}^*({\bf w}) = (k-1) {\bf w}^T {\bf w} + 
[{\bf y}^o - \bar{\bf y}^b - {\bf Y}^b {\bf w}]^T {\bf R}^{-1}
[{\bf y}^o - \bar{\bf y}^b - {\bf Y}^b {\bf w}].
\end{equation}
This cost function is in the form of the Kalman filter cost function
(\ref{linkalcost}), using the background mean $\bar{\bf w}^b = {\bf
0}$ and background covariance $\tilde{\bf P}^b = (k-1)^{-1}
{\bf I}$, with ${\bf Y}^b$ playing the role of the observation
operator.  The analogues of the analysis equations (\ref{anmean'}) and
(\ref{ancov'}) are then
\begin{equation} \label{anmean2}
\bar{\bf w}^a = \tilde{\bf P}^a ({\bf Y}^b)^T {\bf R}^{-1} ({\bf y}^o -
\bar{\bf y}^b),
\end{equation}
\begin{equation} \label{ancov2}
\tilde{\bf P}^a = [(k-1) {\bf I} + ({\bf Y}^b)^T {\bf R}^{-1} {\bf
Y}^b]^{-1}.
\end{equation}

In model space, the analysis mean and covariance are then
\begin{equation} \label{anmean2'}
\bar{\bf x}^a = \bar{\bf x}^b + {\bf X}^b \bar{\bf w}^a,
\end{equation}
\begin{equation} \label{ancov2'}
{\bf P}^a = {\bf X}^b \tilde{\bf P}^a ({\bf X}^b)^T.
\end{equation}
To initialize the ensemble forecast that will produce the
background for the next analysis, we must choose an analysis ensemble
whose sample mean and covariance are equal to $\bar{\bf x}^a$ and
${\bf P}^a$.  As mentioned above, this amounts to choosing a matrix
${\bf X}^a$ so that the sum of its columns is zero and (\ref{enscov})
holds.  Then one can form the analysis ensemble by adding $\bar{\bf
x}^a$ to each of the columns of ${\bf X}^a$.

\paragraph{Symmetric square root.}
Our choice of analysis ensemble is described by ${\bf X}^a = {\bf X}^b
{\bf W}^a$, where
\begin{equation} \label{sqrt}
{\bf W}^a = [(k-1) \tilde{\bf P}^a]^{1/2}
\end{equation}
and by the $1/2$ power of a symmetric matrix we mean its symmetric
square root.  Then $\tilde{\bf P}^a = (k-1)^{-1} {\bf W}^a ({\bf
W}^a)^T$, and (\ref{enscov}) follows from (\ref{ancov2'}).  The use of
the symmetric square root to determine ${\bf W}^a$ from $\tilde{\bf
P}^a$ (as compared to, for example, a Cholesky factorization, or the
choice described in \cite{BEM}), is important for two main reasons.
First, as we will see below, it ensures that the sum of the columns of
${\bf X}^a$ is zero, so that the analysis ensemble has the correct
sample mean (this is also shown for the symmetric square root in
\cite{WBJ}).  Second, it ensures that ${\bf W}^a$ depends continuously
on $\tilde{\bf P}^a$; while this may be a desirable property in
general, it is crucial in a local analysis scheme, so that neighboring
grid points with slightly different matrices $\tilde{\bf P}^a$ do not
yield very different analysis ensembles.
Another potentially desirable property of the symmetric square root is
that it minimizes the (mean-square) distance between ${\bf W}^a$ and
the identity matrix, so that the analysis ensemble perturbations are
in this sense as close as possible to the background ensemble
perturbations subject to the constraint on their sample covariance
\cite{Ott1,Ott2}.

To see that the sum of the columns of ${\bf X}^a$ is zero, we express
this condition as ${\bf X}^a {\bf v} = {\bf 0}$, where ${\bf v}$ is a
column vector of $k$ ones: ${\bf v} = (1, 1, \ldots, 1)^T$.  Notice
that by (\ref{ancov2}), ${\bf v}$ is an eigenvector of
$\tilde{\bf P}^a$ with eigenvalue $(k-1)^{-1}$:
$$
(\tilde{\bf P}^a)^{-1} {\bf v} = [(k-1) {\bf I} + ({\bf Y}^b)^T {\bf
R}^{-1} {\bf Y}^b] {\bf v} = (k-1) {\bf v},
$$
because the sum of the columns of ${\bf Y}^b$ is zero.  Then by
(\ref{sqrt}), ${\bf v}$ is also an eigenvector of ${\bf W}^a$ with
eigenvalue $1$.  Since the sum of the columns of ${\bf X}^b$ is zero,
${\bf X}^a {\bf v} = {\bf X}^b {\bf W}^a {\bf v} = {\bf X}^b {\bf v} =
{\bf 0}$ as desired.

Finally, notice that we can form the analysis ensemble first in
$\tilde{S}$ by adding $\bar{\bf w}^a$ to each of the columns of ${\bf
W}^a$; let $\{{\bf w}^{a(i)}\}$ be the columns of the resulting
matrix.  These ``weight'' vectors specify what linear combinations of
the background ensemble perturbations to add to the background mean
to obtain the analysis ensemble in model space:
\begin{equation} \label{update}
{\bf x}^{a(i)} = \bar{\bf x}^b + {\bf X}^b {\bf w}^{a(i)}.
\end{equation}

\paragraph{Local implementation.}
Notice that once the background ensemble has been used to form
$\bar{\bf y}^b$ and ${\bf Y}^b$, it is no longer needed in the
analysis, except in (\ref{update}) to translate the results from
$\tilde{S}$ to model space.  This point is useful to keep in mind when
implementing a local filter that computes a separate analysis for each
model grid point.  In principle, one should form a global background
observation ensemble ${\bf y}_{[g]}^{b(i)}$ from the global background
vectors, though in practice this can be done locally when the global
observation operator $H_{[g]}$ uses local interpolation.  After the
background observation ensemble is formed, the analyses at different
grid points are completely independent of each other and can be
computed in parallel.  The observations chosen for a given local
analysis dictate which coordinates of ${\bf y}_{[g]}^{b(i)}$ are used
to form the local background observation ensemble ${\bf y}^{b(i)}$,
and the analysis in $\tilde{S}$ produces the local weight vectors
$\{{\bf w}^{a(i)}\}$.  Computing the analysis ensemble $\{{\bf
x}^{a(i)}\}$ for the analysis grid point using (\ref{update}) then
requires only using the background model states at that grid point.

As long as the sets of observations used for a pair of neighboring
grid points overlap heavily, the local weight vectors $\{{\bf
w}^{a(i)}\}$ for the two grid points are similar.  In a region over
which the weight vectors do not change much, the analysis ensemble
members are approximately linear combinations of the
background ensemble members, and thus should represent reasonably
``physical'' initial conditions for the forecast model.  However, if
the model requires of its initial conditions high-order smoothness
and/or precise conformance to an conservation law, it may be necessary
to post-process the analysis ensemble members to smooth them and/or
project them onto the manifold determined by the conserved quantities
before using them as initial conditions (this procedure is often
called ``balancing'' in geophysical data assimilation).

In other localization approaches \cite{HM2,HWS,WH}, the influence of an
observation at a particular point on the analysis at a particular model
grid point decays smoothly to zero as the distance between the two
points increases.  A similar effect can be achieved here by
multiplying the entries in the inverse observation error
covariance matrix ${\bf R}^{-1}$ by a factor that decays from one to
zero as the distance of the observations from the analysis grid point
increases.  This ``smoothed localization'' corresponds to gradually
increasing the uncertainty assigned to the observations until beyond a
certain distance they have infinite uncertainty and therefore no
influence on the analysis.

\paragraph{Covariance inflation.}
In practice, an ensemble Kalman filter that adheres strictly to the
Kalman filter equations (\ref{anmean'}) and (\ref{ancov'}) may
fail to synchronize with the ``true'' system trajectory that produces
the observations.  One reason for this is model error, but even with a
perfect model, the filter tends to underestimate the uncertainty in
its state estimate \cite{WH}.  Regardless of the cause, underestimating
the uncertainty leads to overconfidence in the background state
estimate, and, hence, the analysis underweights the observations.
If the discrepancy becomes too large over time, the
observations are essentially ignored by the analysis, and the dynamics
of the data assimilation system become decoupled from the truth.

Generally this tendency is countered by an {\it ad hoc\/} procedure (with at
least one tunable parameter) that inflates either the background
covariance or the analysis covariance during each data assimilation
cycle.  (Doing this is analogous to adding a model error covariance term
to the right side of (\ref{prop2}), as is usually done in the Kalman
filter.)  One ``hybrid'' approach adds a multiple of the background covariance
matrix ${\bf B}$ from the 3D-Var method to the ensemble background covariance
prior to the analysis \cite{HS}.  ``Multiplicative inflation''
\cite{AA,HWS} instead multiplies the background covariance matrix (or
equivalently, the perturbations of the
background ensemble members from their mean) by a constant factor
larger than one.  ``Additive inflation'' adds a small multiple of the
identity matrix to either the background covariance or the analysis
covariance during each cycle \cite{Ott1,Ott2}.  Finally, if one chooses the
analysis ensemble in such a way that each member has a corresponding
member of the background ensemble, then one can inflate the analysis
ensemble by ``relaxation'' toward the background ensemble: replacing
each analysis perturbation from the mean by a weighted average of
itself and the corresponding background perturbation \cite{ZSS}.

Within the framework described in this article, the hybrid approach is
not feasible because it requires the analysis to consider uncertainty
outside the space spanned by the background ensemble.  However, once
the analysis ensemble is formed, one could develop a means of
inflating it in directions (derived from the 3D-Var background
covariance matrix ${\bf B}$ or otherwise) outside the ensemble space
so that uncertainty in these directions is reflected in the background
ensemble at the next analysis step.  Additive inflation is feasible,
but requires substantial additional computation to determine
the adjustment necessary in the $k$-dimensional space $\tilde{S}$ that
corresponds to adding a multiple of the identity matrix to the model
space covariance ${\bf P}^b$ or ${\bf P}^a$.  Relaxation is simple to
implement, and is most efficiently done in $\tilde{S}$ by replacing
${\bf W}^a$ with a weighted average of it and the identity matrix.

Multiplicative inflation can be performed most easily on the analysis
ensemble by multiplying ${\bf W}^a$ by an appropriate factor (namely
$\sqrt{\rho}$, if one wants to multiply the analysis covariance by $\rho$).
To perform multiplicative inflation on the background ensemble
instead, one can multiply ${\bf X}^b$ by such a
factor, and adjust the background ensemble $\{{\bf x}^{b(i)}\}$
accordingly before applying the observation operator $H$ to form the
background observation ensemble $\{{\bf y}^{b(i)}\}$.  A more
efficient approach, which is equivalent if $H$ is linear,
is simply to replace $(k-1) {\bf I}$ by $(k-1)
{\bf I}/\rho$ in (\ref{ancov2}), since $(k-1) {\bf I}$ is the inverse
of the background covariance matrix $\tilde{\bf P}^b$ in the
$k$-dimensional space $\tilde{S}$.  One can check that this has the
same effect on the analysis mean $\bar{\bf x}^a$ and covariance ${\bf
P}^a$ as multiplying ${\bf X}^b$ and ${\bf Y}^b$ by $\sqrt{\rho}$.
If $\rho$ is close to one, this is a good approximation to inflating
the background ensemble before applying the observation operator
even when this operator is nonlinear.

Multiplicative inflation of the background covariance can be thought
of as applying a discount factor to the influence of past observations
on the current analysis.  Since this discount factor is applied during
each analysis, the cumulative effect is that the influence of an
observation on future analyses decays exponentially with time.  The
inflation factor determines the time scale over which observations
have a significant influence on the analysis.  Other methods of
covariance inflation have a similar effect, causing observations from
sufficiently far in the past essentially to be ignored.  Thus,
covariance inflation localizes the analysis in time.  This effect is
especially desirable in the presence of model error, because then the
model can only reliably propagate information provided by the
observations for a limited period of time.

\section{Efficient Computation of the Analysis} \label{cookbook}

Here is a step-by-step description of how to perform the analysis
described in the previous section, designed for efficiency both in
ease of implementation and in the amount of computation and memory
usage.  Of course there are some trade-offs between these objectives,
so in each step we first describe the simplest approach and then in
some cases mention alternate approaches and possible gains in
computational efficiency.  We also give a rough accounting of the
computational complexity of each step, and at the end discuss the
overall computational complexity.  After that, we describe an approach
that in some cases will produce a significantly faster analysis, at
the expense of more memory usage and more difficult implementation, by
reorganizing some of the steps.  As before, we use ``grid point''
in this section to mean a spatial location in the forecast model,
whether or not the model is actually based on a grid geometry; we
use ``array'' to mean a vector or matrix.  The use of ``columns'' and
``rows'' below is for exposition only; one should of course store
arrays in whatever manner is most efficient for one's computing
environment.

The inputs to the analysis are a background ensemble of
$m_{[g]}$-dimensional model state vectors $\{{\bf x}_{[g]}^{b(i)}: i =
1,2,\ldots,k\}$, a function $H_{[g]}$ from the $m_{[g]}$-dimensional
model space to the $\ell_{[g]}$-dimensional observation space, an
$\ell_{[g]}$-dimensional vector ${\bf y}_{[g]}^o$ of observations, and
an $\ell_{[g]} \times \ell_{[g]}$ observation error covariance matrix
${\bf R}_{[g]}$.  The subscript $g$ here signifies that these inputs
reflect the global model state and all available observations, from
which a local subset should be chosen for each local analysis.  How to
choose which observations to use is entirely up to the user of this
method, but a reasonable general approach is to choose those
observations made within a certain distance of the grid point at which
one is doing the local analysis and determine empirically which value
of the cutoff distance produces the ``best'' results.  If one deems
localization to be unnecessary in a particular application, then one
can ignore the distinction between local and global, and skip
Steps~\ref{s3b} and \ref{s9} below.

Steps~\ref{s2a} and \ref{s1a} are essentially global operations, but
may be done locally in a parallel implementation.
Steps~\ref{s3b}--\ref{s7} should be performed separately for
each local analysis (generally this means for each grid point, but see
the parenthetical comment at the end of Step~\ref{s3b}).
Step~\ref{s9} simply combines the results of the local analyses to
form a global analysis ensemble $\{{\bf x}_{[g]}^{a(i)}\}$, which is the
final output of the analysis.

\begin{enumerate}

\item \label{s2a}
{\em Apply $H_{[g]}$ to each ${\bf x}_{[g]}^{b(i)}$ to form the global
background observation ensemble $\{{\bf y}_{[g]}^{b(i)}\}$, and
average the latter vectors to get the $\ell_{[g]}$-dimensional column
vector $\bar{\bf y}_{[g]}^b$.  Subtract this vector from each $\{{\bf
y}_{[g]}^{b(i)}\}$ to form the columns of the $\ell_{[g]} \times k$
matrix ${\bf Y}_{[g]}^b$.}  (This subtraction can be done ``in
place'', since the vectors $\{{\bf y}_{[g]}^{b(i)}\}$ are no longer
needed.)  This requires $k$ applications of $H$, plus $2k\ell_{[g]}$
(floating-point) operations.  If $H$ is an interpolation operator that
requires only a few model variables to compute each observation
variable, then the total number of operations for this step is
proportional to $k\ell_{[g]}$ times the average number of model
variables required to compute each scalar observation.

\item \label{s1a}
{\em Average the vectors $\{{\bf x}_{[g]}^{b(i)}\}$ to get the
$m_{[g]}$-dimensional vector $\bar{\bf x}_{[g]}^b$, and subtract this
vector from each ${\bf x}_{[g]}^{b(i)}$ to form the columns of the
$m_{[g]} \times k$ matrix ${\bf X}_{[g]}^b$.}  (Again the subtraction
can be done ``in place''; the vectors $\{{\bf x}_{[g]}^{b(i)}\}$ are
no longer needed).  This step requires a total of $2km_{[g]}$
operations.  (If $H$ is linear, one can equivalently perform
Step~\ref{s1a} before Step~\ref{s2a}, and obtain $\bar{\bf y}_{[g]}^b$
and ${\bf Y}_{[g]}^b$ by applying $H$ to $\bar{\bf x}_{[g]}^b$ and
${\bf X}_{[g]}^b$.)

\item \label{s3b}
This step selects the necessary data for a given grid point (whether
it is better to form the local arrays described below explicitly or
select them later as needed from the global arrays depends on one's
implementation).  {\em Select the rows of $\bar{\bf x}_{[g]}^b$ and
${\bf X}_{[g]}^b$ corresponding to the given grid point, forming their
local counterparts: the $m$-dimensional vector $\bar{\bf x}^b$ and the
$m \times k$ matrix ${\bf X}^b$, which will be used in Step~\ref{s7}.
Likewise, select the rows of $\bar{\bf y}_{[g]}^b$ and ${\bf
Y}_{[g]}^b$ corresponding to the observations chosen for the analysis
at the given grid point, forming the $\ell$-dimensional vector
$\bar{\bf y}^b$ and the $\ell \times k$ matrix ${\bf Y}^b$.  Select
the corresponding rows of ${\bf y}^o_{[g]}$ and rows and columns of
${\bf R}_{[g]}$ to form the $\ell$-dimensional vector ${\bf y}^o$ and
the $\ell \times \ell$ matrix ${\bf R}$.}  (For a high-resolution
model, it may be reasonable to use the same set of observations for
multiple grid points, in which case one should select here the rows of
${\bf X}_{[g]}^b$ and $\bar{\bf x}_{[g]}^b$ corresponding to all of
these grid points.)

\item \label{s3a}
{\em Compute the $k \times \ell$ matrix ${\bf C} = ({\bf Y}^b)^T {\bf
R}^{-1}$.}  If desired, one can multiply entries of ${\bf R}^{-1}$ or
${\bf C}$ corresponding to a given observation by a factor less than
one to decrease (or greater than one to increase) its influence on the
analysis.  (For example, one can use a multiplier that depends on
distance from the analysis grid point to discount observations near
the edge of the local region from which they are selected; this will
smooth the spatial influence of observations, as described in
Section~\ref{letkf} under ``Local Implementation''.)
Since this is the only step in which ${\bf R}$ is used, it
may be most efficient to compute ${\bf C}$ by solving the linear
system ${\bf R} {\bf C}^T = {\bf Y}^b$ rather than inverting ${\bf
R}$.  In some applications, ${\bf R}$ may be diagonal, but in others
${\bf R}$ will be block diagonal with each block representing a group
of correlated observations.  As long as the size of each block is
relatively small, inverting ${\bf R}$ or solving the linear system
above will not be computationally expensive.  Furthermore, many or all
of the blocks that make up ${\bf R}$ may be unchanged from one
analysis time to the next, so that their inverses need not be
recomputed each time.  Based on these considerations, the number of
operations required (at each grid point) for this step in a typical
application should be proportional to $k\ell$, multiplied by a factor
related to the typical block size of ${\bf R}$.

\item \label{s4}
{\em Compute the $k \times k$ matrix $\tilde{\bf P}^a = \left[(k-1){\bf
I}/\rho + {\bf C}{\bf Y}^b\right]^{-1}$, as in (\ref{ancov2}).}  Here
$\rho > 1$ is a multiplicative covariance inflation factor, as
described at the end of the previous section.  Though trying some of
the other approaches described there may be fruitful, a reasonable
general approach is to start with $\rho > 1$ and increase it gradually
until one finds a value that is optimal according to some measure of
analysis quality.  Multiplying ${\bf C}$ and ${\bf Y}^b$ requires
less than $2k^2\ell$ operations, while the number of operations needed
to invert the $k \times k$ matrix is proportional to $k^3$.

\item \label{s5}
{\em Compute the $k \times k$ matrix ${\bf W}^a = [(k-1)\tilde{\bf
P}^a]^{1/2}$, as in (\ref{sqrt}).}  Again the number of operations
required is proportional to $k^3$; it may be most efficient to compute
the eigenvalues and eigenvectors of $\left[(k-1){\bf I}/\rho + {\bf
C}{\bf Y}^b\right]$ in the previous step and then use them to compute
both $\tilde{\bf P}^a$ and ${\bf W}^a$.

\item \label{s6}
{\em Compute the $k$-dimensional vector $\bar{\bf w}^a = \tilde{\bf
P}^a{\bf C}({\bf y}^o - \bar{\bf y}^b)$, as in (\ref{anmean2}), and
add it to each column of ${\bf W}^a$, forming a $k \times k$ matrix
whose columns are the analysis vectors $\{{\bf w}^{a(i)}\}$.}
Computing the formula for $\bar{\bf w}^a$ from right-to-left, the total
number of operations required for this step is less than $3k(\ell +
k)$.

\item \label{s7}
{\em Multiply ${\bf X}^b$ by each ${\bf w}^{a(i)}$ and add $\bar{\bf
x}^b$ to get the analysis ensemble members $\{{\bf x}^{a(i)}\}$ at the
analysis grid point, as in (\ref{update}).}  This requires $2k^2m$
operations.

\item \label{s9}
{\em After performing Steps~\ref{s3b}--\ref{s7} for each grid point,
the outputs of Step~\ref{s7} form the global analysis ensemble $\{{\bf
x}_{[g]}^{a(i)}\}$.}

\end{enumerate}

We now summarize the overall computational complexity of the algorithm
described above.  If $p$ is the number local analyses performed (equal
to the number of grid points in the most basic approach), then notice
that $pm = m_{[g]}$, while $p\bar{\ell} = q\ell_{[g]}$, where $\bar{\ell}$ is
the average number of observations used in a local analysis and $q$ is
the average number of local analyses in which a particular observation
is used.  If $\bar{\ell}$ is large compared to $k$ and $m$, then the
most computationally expensive step is either Step~\ref{s4}, requiring
approximately $2k^2 p\bar{\ell} = 2k^2 q\ell_{[g]}$ operations over all
the local analyses, or Step~\ref{s3a}, whose overall number of
operations is proportional to $k p\bar{\ell} = k q\ell_{[g]}$, but with a
proportionality constant dependent on the correlation structure of
${\bf R}_{[g]}$.  In any case, as long as the typical number of correlated
observations in a block of ${\bf R}_{[g]}$ remains constant, the overall
computation time grows at most linearly with the total number $\ell_{[g]}$
of observations.  It also grows at most linearly with the total number
$m_{[g]}$ of model variables; if $m_{[g]}$ is large enough compared to
$\ell{[g]}$, then the most expensive step is Step~\ref{s7}, with $2k^2
m_{[g]}$ overall operations.  The terms in the computation time that
grow with the number of observations or number of model variables are
at most quadratic in the number $k$ of ensemble members.  However, for
a sufficiently large ensemble, the matrix operations in Steps~\ref{s4}
and \ref{s5} that take of order $k^3$ operations per local analysis,
or $k^3p$ operations overall, will become significant.

In Section~\ref{results}, we present some numerical results for which
we find the computation time indeed grows roughly quadratically with
$k$, linearly with $q$, and sublinearly with $\ell_{[g]}$.

\paragraph{Batch processing of observations.}
Some of the steps above have a $q$-fold redundancy, in that
computations involving a given observation are repeated over an
average of $q$ different local analyses.  For a general observation
error covariance matrix ${\bf R}_{[g]}$ this redundancy may be
unavoidable, but it can be avoided as described below if the global
observations can be partitioned into local groups (or ``batches'')
numbered $1, 2, \ldots$ that meet the following conditions.  First,
all of the observations in a given batch must be used in the exact
same subset of the local analyses.  Second, observations in different
batches must have uncorrelated errors, so that each batch $j$
corresponds to a block ${\bf R}_j$ in a block diagonal decomposition
of ${\bf R}_{[g]}$.  (These conditions can always met if ${\bf
R}_{[g]}$ is diagonal, by making each batch consist of a single
observation.  However, as explained below, for efficiency one should
make the batches as large as possible while still meeting the first
condition.)  Then at Step~\ref{s3b}, instead of selecting
(overlapping) submatrices of $\bar{\bf y}_{[g]}^b$, ${\bf Y}_{[g]}^b$,
${\bf y}_{[g]}^o$, and ${\bf R}_{[g]}$, for each grid point, let
$\bar{\bf y}_j^b$, ${\bf Y}_j^b$, ${\bf y}_j^o$, represent the rows
corresponding to the observations in batch $j$, and do the following
for each batch.  Compute and store the $k \times k$ matrix ${\bf C}_j
{\bf Y}_j^b$ and the $k$-dimensional vector ${\bf C}_j ({\bf y}_j^o -
\bar{\bf y}_j^b)$, where ${\bf C}_j = ({\bf Y}_j^b)^T {\bf R}_j^{-1}$
as in Step~\ref{s3a}.  (This can be done separately for each batch, in
parallel, and the total number of operations is roughly $2k^2
\ell_{[g]}$.)  Then do Steps~\ref{s4}--\ref{s7} separately for each
local analysis; when ${\bf C}{\bf Y}^b$ and ${\bf C} ({\bf y}^o -
\bar{\bf y}^b)$ are required in Steps~\ref{s4} and \ref{s6}, compute
them by summing the corresponding arrays ${\bf C}_j {\bf Y}_j^b$ and
${\bf C}_j ({\bf y}_j^o - \bar{\bf y}_j^b)$ over the batches $j$ of
observations that are used in the local analysis.  To avoid redundant
addition in these steps, batches that are used in exactly the same
subset of the local analyses should be combined into a single batch.
The total number of operations required by the summations over batches
roughly $k^2 ps$, where $s$ is the average number of batches used in
each local analysis.  Both this and the $2k^2 \ell_{[g]}$ operations
described before are smaller than the roughly $2k^2 p\bar{\ell} = 2k^2
q\ell_{[g]}$ operations they combine to replace.

This approach has similarities with the ``sequential'' approach of
\cite{HM2} and \cite{WH}, in which observations are divided into
uncorrelated batches and a separate analysis is done for each batch;
the analysis is done in the observation space whose dimension is the
number of observations in a batch.  However, in the sequential
approach, the analysis ensemble for one batch of observations is
used as the background ensemble for the next batch of observations.
Since batches with disjoint local regions of influence can be analyzed
separately, some parallelization is possible, though the LETKF
approach described above is more easily distributed over a large
number of processors.  For a serial implementation, either approach
may be faster depending on the application and the ensemble size.

\section{Asynchronous Observations: 4D-LETKF} \label{4dletkf}

In theory, one can perform a new analysis each time new observations
are made.  In practice, this is a good approach if observations are
made at regular and sufficiently infrequent time intervals.  However,
in many applications, such as weather forecasting, observations are
much too frequent for this approach.  Imagine, for example, a 6-hour
interval between analyses, like at the National Weather Service.
Since weather can change significantly over such a time interval, it
is important to consider observations taken at intermediate times in a
more sophisticated manner than to pretend that they occur at the
analysis time (or to simply ignore them).  Operational versions of
3D-Var and 4D-Var (see Section~\ref{nonlin}) do take into account the
timing of the observations, and one of the primary strengths of 4D-Var
is that it does so in a precise manner, by considering which forecast
model trajectory best fits the observations over a given time interval
(together with assumed background statistics at the start of this
interval).

We have seen that the analysis step in an ensemble Kalman filter
considers model states that are linear combinations of the background
ensemble states at the analysis time, and compares these model states
to observations taken at the analysis time.  Similarly, we can
consider approximate model trajectories that are linear combinations
of the background ensemble trajectories over an interval of time, and
compare these approximate trajectories with the observations taken
over that time interval.  Instead of asking which model trajectory
best fits the observations, we ask which linear combination of the
background ensemble trajectories best fits the observations.  As
before, this is relatively a low-dimensional optimization problem that
is much more computationally tractable than the full nonlinear
problem.

This approach is similar to that of an ensemble Kalman smoother
\cite{EV,Eve2}, but over a much shorter time interval.  As compared to
a ``filter'', which estimates the state of a system at time $t$ using
observations made up to time $t$, a ``smoother'' estimates the system
state at time $t$ using observations made before and after time $t$.
Over a long time interval, one must generally take a more
sophisticated approach to smoothing than to simply consider linear
combinations of an ensemble of trajectories generated over the entire
interval, both because the trajectories may diverge enough that linear
combinations of them will not approximate model trajectories, and
because in the presence of model error there may be no model
trajectory that fits the observations over the entire interval.  Over
a sufficiently short time interval however, the approximation of true system
trajectories by linear combinations of model trajectories with similar
initial conditions is quite reasonable.

While this approach to assimilating asynchronous observations is
suitable for any ensemble Kalman filter \cite{Hun}, it is particularly
simple to implement in the LETKF framework.  We call this extension
4D-LETKF; see \cite{HH2} for an alternate derivation of this algorithm.

To be more concrete, suppose that we have observations $(\tau_j,
{\bf y}^o_{\tau_j})$ taken at various times $\tau_j$ since the previous
analysis.  Let $H_{\tau_j}$ be the observation operator for time
$\tau_j$ and let ${\bf R}_{\tau_j}$ be the error covariance matrix for
these observations.  In Section ~\ref{letkf}, we mapped a vector ${\bf
w}$ in the $k$-dimensional space $\tilde{S}$ into observation space
using the formula $\bar{\bf y}^b + {\bf Y}^b {\bf w}$, where the
background observation mean $\bar{\bf y}^b$ and perturbation matrix
${\bf Y}^b$ were formed by applying the observation operator $H$ to
the background ensemble at the analysis time.  So now, for each time
$\tau_j$, we apply $H_{\tau_j}$ to the background ensemble at time
$\tau_j$, calling the mean of the resulting vectors $\bar{\bf
y}^b_{\tau_j}$ and forming their differences from the mean into the
matrix ${\bf Y}^b_{\tau_j}$.

We now form a combined observation vector ${\bf y}^o$ by concatenating
(vertically) the (column) vectors ${\bf y}^o_{\tau_j}$, and similarly
by vertical concatenation of the vectors $\bar{\bf y}^b_{\tau_j}$ and
matrices ${\bf Y}^b_{\tau_j}$ respectively, we form the combined
background observation mean $\bar{\bf y}^b$ and perturbation matrix
${\bf Y}^b$.  We form the corresponding error covariance matrix ${\bf
R}$ as a block diagonal matrix with blocks ${\bf R}_{\tau_j}$
(this assumes that observations taken at different times have
uncorrelated errors, though such correlations if present could be
included in ${\bf R}$).

Given this notation, we can then use the same analysis equations as in
the previous sections, which are based on minimizing the cost function
$\tilde{J}^*$ given by (\ref{lincost}).  (We could instead write
down the appropriate analogue to (\ref{tildecost}) and minimize the
resulting nonlinear cost function $\tilde{J}$; this would be no harder
in principle than in the case of synchronous observations.)  Referring to
Section~\ref{cookbook}, the only change is in Step~\ref{s2a}, which
one should perform for each observation time $\tau_j$ (using the
background ensemble and observation operator for that time) and then
concatenate the results as described above.  Step~\ref{s1a} still only
needs to be done at the analysis time, since its output is used
only in Step~\ref{s7} to form the analysis ensemble in model space.
All of the intermediate steps work exactly the same, in terms of the
output of Step~\ref{s2a}.

In practice, the model will be integrated with a discrete time step
that in general will not coincide with the observation times
$\tau_j$.  One should either interpolate the background ensemble
trajectories to the observation times, or simply round the observation
times off to the nearest model integration time.  In either case, one
must either store the background ensemble trajectories until the
analysis time, or perform Step~\ref{s2a} of Section~\ref{cookbook}
during the model integration and store its results.  The latter
approach will require less storage if the number of observations per
model integration time step is less than the number of model
variables.

One can perform localization in the same manner as with synchronous
observations, but it may be advantageous to take into account the
timing of the observations when deciding which of them to use in a
given local analysis.  For example, due to spatial propagation in the
model dynamics, one may wish to include earlier observations from a
greater distance than later observations.  On the other hand, earlier
observations may be less useful than observations closer to the
analysis time due to model error; it may help then to decrease the
influence of the earlier observations as described in Step~\ref{s3a}
of Section~\ref{cookbook}.

\section{Numerical Experiments with Real Atmospheric Observations}
\label{results}

We have implemented the 4D-LETKF algorithm, as described in
Sections~\ref{cookbook} and \ref{4dletkf}, with the operational Global
Forecast System (GFS) model \cite{GFS} of the U.S. National Centers
for Environmental Prediction (NCEP).  This model is used (currently at
higher resolution than we describe below) for National Weather Service
forecasts.  Previously we have published results using this model with
the LEKF algorithm of \cite{Ott1,Ott2}, in a perfect model scenario
(with simulated observations) \cite{Szu}.  Using the same parameters
for the LETKF algorithm, we have obtained results very similar to
those in \cite{Szu}, which we do not repeat here; with LETKF, the data
assimilation steps run $3$ to $5$ times as fast as with LEKF.

Here, we present some preliminary results obtained using the 4D-LETKF
algorithm with the same model and real atmospheric observations
collected in January and February 2004, and compare them with results
from the NCEP Spectral Statistical Interpolation (SSI) \cite{SSI}, a
state-of-the-art implementation of 3D-Var.  Further results will
appear in a future publication.  Our data set includes all
operationally assimilated observations except for satellite radiances
and measurements of atmospheric humidity.  Observations include
vertical sounding profiles of temperature and wind by weather
balloons, surface pressure observations by land and sea stations,
temperature and wind reports by commercial aircraft, and wind vectors
derived from satellite based observation of clouds.

For all of the results below, we assimilate observations every 6
hours, and we use a model resolution of T62 (a $192 \times 94$
longitude-latitude grid) with 28 vertical levels,
for a total of about $500,000$ points.  In our
4D-LETKF implementation, for each grid point, we selected observations
from within a $h \times h \times v$ subset of the model grid, centered
at the analysis grid point, with the vertical height $v$ varying
(depending on the vertical level) from $1$ to
$7$ grid points as in \cite{Szu}, and the horizontal width $h$ held
constant for each experiment at either $5$ or $7$ grid points.  The
number of ensemble members $k$ we use in each experiment is either
$40$ or $80$.  In all 4D-LETKF experiments we used a
spatially-dependent multiplicative covariance inflation factor
$\rho$, which we taper from $1.15$ at the surface to $1.1$ at the top
of the model atmosphere in the Southern Hemisphere, and from $1.25$ at
the surface to $1.15$ at the top in the Northern Hemisphere (between
$30^\circ$S and $30^\circ$N latitudes, we linearly interpolate between
these values).

\subsection{Analysis Quality}

In this section, we compare the analyses from our 4D-LETKF
implementation, using $k = 40$ ensemble members and a $7 \times 7$
horizontal grid ($h = 7$) for each local region, and from the NCEP
SSI, using the same model resolution and the same observational data
set for its global analysis (we call this the ``benchmark'' SSI
analysis).  To estimate the analysis error for a given state variable,
we compute the spatial RMS difference between its analysis and the
operational high-resolution SSI analysis computed by NCEP (we call
this the ``verification'' SSI analysis).  While this verification
technique favors the benchmark SSI analysis, which is obtained with
the same data assimilation method, it can provide useful information
in regions where the 4D-LETKF and benchmark SSI analyses exclude a
large portion of the observations assimilated into the verifying SSI
analysis.  Such a region is the Southern Hemisphere, where satellite
radiances are known to have a strong positive impact on the quality of
the analysis.

We initialize 4D-LETKF with a random ensemble of physically plausible
global states at midnight on 1 January 2004.  Specifically, we take
each initial ensemble member from an operational NCEP analysis
at a randomly chosen time between 15 January and 31 March 2004.
The 4D-LETKF analyses start to synchronize with the observations after
a few days.  To exclude from our comparison the transient error due to
initialization of 4D-LETKF, we average all estimated errors over the
month of February 2004 only.

Figure~1 shows the estimated analysis error of each method for
temperature in the Southern Hemisphere extratropics ($20^\circ$S to
$90^\circ$S latitudes) as a function of atmospheric pressure.  The
4D-LETKF analysis is more accurate than the benchmark SSI at all
except near the surface, where the two methods are quite
similar in accuracy.  The advantage of the 4D-LETKF analysis is
especially large in the upper atmosphere, where observations are
extremely sparse.  Figure~2 makes the same comparison between the
48-hour forecasts generated from the respective analyses, again
verified against the operational high-resolution SSI analysis at the
appropriate time.  We see that the 4D-LETKF forecasts are also more
accurate than those from the background SSI analysis, especially in
the upper atmosphere.

\begin{figure}[ht]
\begin{center}
\includegraphics[height=3in]{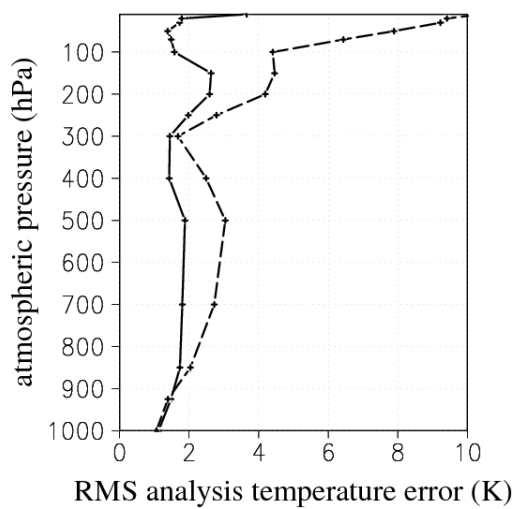}
\end{center}
\caption{Vertical profile of the estimated analysis temperature error,
in degrees Kelvin, for the 4D-LETKF (solid) and benchmark SSI (dashed)
analyses in the Southern Hemisphere extratropics.  The atmospheric
height is indicated by pressure, in hectopascal.  The estimate of the
error is obtained by calculating the
root-mean-square difference between each analysis and the verifying
SSI analysis for latitudes between $20^\circ$S and $90^\circ$S and
averaging over all the analysis times in February 2004.}
\end{figure}

\begin{figure}[ht]
\begin{center}
\includegraphics[height=3in]{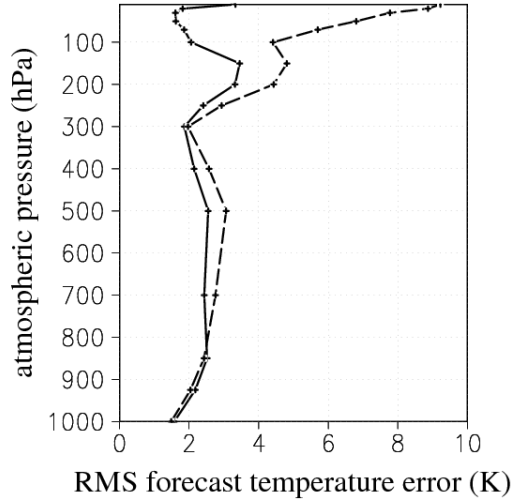}
\end{center}
\caption{Estimated 48-hour forecast
temperature error versus atmospheric pressure for the 4D-LETKF
(solid) and benchmark SSI (dashed) methods in the Southern Hemisphere
extratropics.  The estimated forecast error is computed in the same
way as the estimated analysis error shown in Figure~1.}
\end{figure}

For wind in the Southern Hemisphere and temperature and wind in the
Northern Hemisphere, the RMS analysis and forecast errors for the
4D-LETKF and benchmark SSI are more similar to each other, though in
all cases the 4D-LETKF results are significantly better in the highest
part of the atmosphere ($0$ hPa to $100$ hPa).

We emphasize that these results are obtained with modest tuning of the
4D-LETKF parameters, and we expect further significant improvements
from a more thorough exploration of the algorithm's parameter space,
as well as a more sophisticated approach to model error, such as the
adaptive bias-correction technique of \cite{BHKOS}.

Qualitatively similar results with the same model and a similar data
set are reported in \cite{WHWST}, using both an alternate
implementation of the 4D-LETKF algorithm (with a different covariance
inflation approach) and a related method based on the Ensemble
Square-Root Filter of \cite{WH}.  The latter method was slightly more
accurate in the Northern Hemisphere and slightly less accurate in the
Southern Hemisphere, and both methods were more accurate than the
corresponding benchmark SSI, when verified both against the
operational high-resolution SSI analysis and against observations.
See also \cite{HMPBCSH} and \cite{MY} for comparisons of ensemble
Kalman filter results to those of other operational 3D-Var methods,
using forecast models from the Meteorological Service of Canada and the
Japan Meteorological Agency, respectively; the latter article also
implements the LETKF approach.

\subsection{Computational Speed}

In this section, we present and discuss timing results from several
representative analyses of the 4D-LETKF experiment using the GFS
described above, with different numbers of observations.  In addition,
we vary the number of ensemble members ($k = 40$ or $80$) and the
horizontal width of the local region ($h = 5$ or $7$ grid points in
both latitude and longitude).  Though we use a parallel
implementation, we report in Table~1 below the total CPU time used on
a Linux cluster of forty $3.2$ GHz Intel Xeon processors.  The actual
run time is many times faster; with the larger local region ($h = 7$)
the analysis takes about $6$ minutes on our cluster with $k = 40$
ensemble members, and $18$ minutes with $k = 80$.  Thus, the results
shown in Figures~1 and 2 can be obtained in an operational setting
that allots only a few minutes for each analysis.  Furthermore,
because the observations are very nonuniformly distributed spatially,
we expect to be able to reduce the parallel run time considerably by
balancing the load more evenly between processors.  We will report
details of our parallel implementation in a future publication.

Table~1 shows the total CPU time in seconds for $4$ different 4D-LETKF
parameter sets at each of $4$ different analysis times.  Different
numbers of observations are available for each analysis time, with
about $50\%$ more at 1200 GMT than at 0600 GMT.  The computation time
generally grows with the number of observations, though not by as
large a factor.  Referring to the discussion immediately following
Steps~\ref{s2a} to \ref{s9} in Section~\ref{cookbook}, this indicates
that the matrix multiplication portion of Step~\ref{s4} that requires
on the order of $k^2q\ell_{[g]}$ total floating point operations is a
significant component of the computation time, but that other parts of
the computation are significant too.  (Recall that $\ell_{[g]}$ is the
global number of observations and $q$ is the average number of
analyses in which each observation is used, which in this
implementation is roughly the average number of grid points per local
region.)  As $h$ increases from $5$ to $7$, the value of $q$
approximately doubles, and so does the computation time.  And as $k$
increases from $40$ to $80$, the computation time grows by a factor of
$4$ to $5$, indicating that the time is roughly quadratic in $k$ but
suggesting that terms that are cubic in $k$ are becoming significant.

\begin{table}[ht]
\begin{center}
\begin{tabular}{|l|r|r|r|r|}
\hline
analysis time & 0600 GMT & 1800 GMT & 0000 GMT & 1200 GMT \\
\hline
\# observations & 159,947 & 193,877 & 236,168 & 245,850 \\
\hline
\hline
$k=40$, $h=5$ & 945 & 945 & 1244 & 1142 \\
\hline
$k=40$, $h=7$ & 1846 & 2076 & 2105 & 2200 \\
\hline
$k=80$, $h=5$ & 4465 & 4453 & 5124 & 5010 \\
\hline
$k=80$, $h=7$ & 9250 & 10631 & 10463 & 10943 \\
\hline
\end{tabular}
\end{center}
\caption{Total CPU time in seconds (on 3.2GHz Intel Xeon processors)
for various analyses with 4D-LETKF using the GFS with approximately
$500,000$ model grid points.  Columns
represent different analysis times, arranged in increasing order of
the number of observations assimilated.  Rows represent different
values of the ensemble size $k$ and horizontal localization width
$h$.  Notice that even on a single processor, all of these analyses
can be done in less than real time.}
\end{table}

Indeed, examining the CPU time spent in various subroutines on
different processors confirms that most of the time is spent in
Steps~\ref{s4} and \ref{s5}, and that in local analyses where
observations are dense, the matrix multiplication in Step~\ref{s4}
dominates the computation time, while in local analyses where
observations are sparse, the matrix inverse and square root in
Steps~\ref{s4} and \ref{s5} dominate.  We find that the latter
operations take more time in the analyses with the larger local
region; this suggests that the iterate eigenvalue routine we use takes
longer to run in cases when the presence of more observations causes
$\tilde{\bf P}^a$ to be further from a multiple of the identity
matrix.  There is also some computational overhead not accounted for
in Section~\ref{cookbook} whose contribution to the computation time
is not negligible, in particular determining which observations to use
for each local analysis.

Overall, our timing results indicate that with a model and data set of
this size, a substantially larger ensemble size than we currently use
may be problematic, but that our implementation of 4D-LETKF should be
able to assimilate more observations with at most linear growth in the
computation time.  Furthermore, though we do not vary the number of
model variables in Table~1, our examination of the time spent
performing each of the steps from Section~\ref{cookbook} suggests that
we can increase the model resolution significantly without having much
effect on the analysis computation time (though the time spent running
the model would of course increase accordingly).

\section{Summary and Acknowledgments} \label{discuss}

In this article, we have described a general framework for data
assimilation in spatiotemporally chaotic systems using an ensemble
Kalman filter that in its basic version (Section~\ref{cookbook}) is
relatively efficient and simple to implement.  In a particular
application, one may be able to improve accuracy by experimenting with
different approaches to localization (see the discussion in
Sections~\ref{nonlin} and \ref{letkf}), covariance inflation
(see the end of Section~\ref{letkf}), and/or asynchronous observations
(Section~\ref{4dletkf}).  For very large systems and/or when maximum
efficiency is important, one should consider carefully the comments
about implementation in Section~\ref{cookbook} (and at the end of
Section~\ref{4dletkf}, if applicable).  One can also apply this method
to low-dimensional chaotic systems, without using localization.

In Section~\ref{results}, we presented preliminary results for a
relatively straightforward implementation of the LETKF approach with
real atmospheric data and an operational global forecast model.  Our
results demonstrate that this implementation can produce results of
operational quality within a few minutes on a parallel computer of
reasonable size.  The efficiency of the basic algorithm provides many
opportunities to improve the quality of the results with the
variations discussed and referred to in this article.

Many colleagues have contributed to this article in various ways.  We
thank M. Cornick, E. Fertig, J. Harlim, H. Li, J. Liu, T. Miyoshi, and
J. Whitaker for sharing their thoughts and experiences implementing
and testing the LETKF algorithm in a variety of scenarios.  We also
thank C. Bishop, T. Hamill, K. Ide, E. Kalnay, E. Ott, D. Patil, T. Sauer,
J. Yorke, M. Zupanski, and the anonymous reviewers for their generous
input.  This feedback resulted in many improvements to the exposition
in this article and to our implementation of the algorithm.  We thank
Y. Song, Z. Toth, and R. Wobus for providing us with the observations
and benchmark analyses used in Section~\ref{results}, and we thank
G. Gyarmati for developing software to read the observations on our
computers.  This research was supported by grants from NOAA/THORPEX,
the J. S. McDonnell Foundation, and the National Science Foundation
(grant \#ATM034225).  The second author gratefully acknowledges
support from the NSF Interdisciplinary Grants in the Mathematical
Sciences program (grant \#DMS0408012).

\end{document}